\documentclass[12pt,a4paper]{article}
\usepackage{adjustbox}
\usepackage{multirow}
\usepackage{subcaption}
\usepackage{float}
\usepackage{hyperref}
\usepackage{url}
\usepackage{graphicx}
\usepackage{amssymb,amsmath,bm}
\usepackage{comment}
\usepackage{color}
\usepackage{booktabs}

\hypersetup{
   colorlinks,
   menucolor=black,
   linkcolor=black,
   citecolor=red,
   urlcolor=blue
}

\begin{document}

\title{An examination of the alleged privacy threats of confidence-ranked reconstruction of Census microdata}

\author{David S\'anchez$^1$, Najeeb Jebreel$^1$, Krishnamurty Muralidhar$^2$,\\ 
Josep Domingo-Ferrer$^1$, and 
Alberto Blanco-Justicia$^1$\\
{\small $^1$ Universitat Rovira i Virgili}\\
{\small Dept. of Computer Engineering and Mathematics}\\
{\small CYBERCAT-Center for Cybersecurity Research of Catalonia}\\
{\small Av. Pa\"{\i}sos Catalans 26}\\
{\small 43007 Tarragona, Catalonia}\\
{\small \{david.sanchez, najeeb.jebreel,josep.domingo,alberto.blanco\}@urv.cat}\\
{\small $^2$ University of Oklahoma}\\
{\small Price College of Business}\\
{\small Dept. of Marketing and Supply Chain Management}\\
{\small 307 West Brooks, Adams Hall Room 10}\\
{\small Norman, OK 73019, U.S.A.}\\
{\small krishm@ou.edu}
}

\maketitle

\begin{abstract}
The threat of reconstruction attacks has led the U.S. Census Bureau (USCB) to replace in the Decennial Census 2020
the traditional statistical disclosure limitation based on rank swapping with one based on differential privacy (DP),
leading to substantial accuracy loss of released statistics. 
Yet, it has been argued that, if many different reconstructions are compatible with the released statistics,
most of them do not correspond to actual original data, which protects against respondent reidentification. 
Recently, a new attack has been proposed, which incorporates the confidence that a reconstructed record was in the original data.
The alleged risk of disclosure entailed by such confidence-ranked reconstruction has renewed the interest of the USCB to use DP-based solutions.
To forestall a potential accuracy loss in future releases, we show that the proposed reconstruction is neither effective as a reconstruction method
nor conducive to disclosure as claimed by its authors.
Specifically, we report empirical results showing the proposed ranking cannot guide reidentification or attribute disclosure attacks, 
and hence fails to warrant the utility sacrifice entailed by the use of DP to release census statistical data. \\
{\bf Keywords:} U.S. Decennial Census 2020, Statistical Disclosure Limitation, Reconstruction, Differential Privacy, Reidentification.
\end{abstract}

\label{significance}

\section{Introduction}\label{sec1}
The U.S. Decennial Census is the world's most prominent census data release, accounting for more than 330 million people.
The policies implemented by the U.S. Census Bureau (USCB) ---the statistical agency in charge of compiling, managing, and releasing the U.S. Census data---
carry therefore a great influence on the decisions made by many other statistical agencies. 

For the Decennial Census 2020 release, the USCB decided to replace the statistical disclosure limitation (SDL) methodology
they had been using in previous editions (based on data swapping) with a method based on differential privacy (DP).
The reason for such a change was the alleged disclosure risks
resulting from potential record reconstruction attacks on the prior 2010 Census release \cite{abowd2021,hawes2022}.
These works claim it is possible to accurately reconstruct the individual records from the released census statistics
and that this reconstruction entails serious privacy threats to citizens. To prevent accurate reconstruction,
a DP-based method consisting in adding random noise to the released statistics was implemented in the 2020 release \cite{abowd2023}. 

DP is a robust privacy-enhancing method when properly implemented \cite{Domingo2021}, but it also introduces significant logical
inconsistencies and inaccuracies in the protected data due to its random and perturbative nature \cite{santos2020,dove2021}. 
In fact, in many cases the 2020 Census DP-protected data have been manipulated to the point where relationships between variables
are not possible on the ground: there are blocks with more households than household population, blocks with no population in households
yet having occupied house units, or blocks with non-zero population but with no adults \cite{menger2021}.
In this respect, the quality of block-level data in the 2020 release was so poor that the USCB acknowledged that block-level data should
not be used for any meaningful analysis \cite{census2021}.
Given that census data are crucial for research and social decision-making, one can understand the consternation of potential users 
\cite{kenny2021,ruggles2022,hotz2022,schneider2022}. 

Several authors have demonstrated that the claims of disclosure risk ensuing from the reconstruction attacks in \cite{abowd2021} and \cite{hawes2022}
were vastly overstated \cite{ruggles2022,Muralidhar2022}.
A major issue is that there are usually a lot of different reconstructions compatible with the released output,
which makes it impossible for the attacker to know which is the good reconstruction (the one corresponding or
closest to the original data) \cite{Muralidhar2022,Muralidhar2023}.

After assessing the situation, the USCB decided to abstain from using DP to protect the American Community Survey
for the foreseeable future since ``it’s also not clear that differential privacy would ultimately be the best option'' \cite{daily2022}.

However, a recent work by Dick \emph{et al.} claims it is possible to rank records reconstructed by an attacker to reflect how
likely they are to belong to the original data \cite{dick2023confidence}.
The authors propose a \emph{confidence-ranked reconstruction attack} (\textbf{CRR}) that reconstructs records from the
published census statistics and ranks them by how frequently they appear in multiple reconstructions.
The authors assert that their ranking can be used by an adversary to conduct a variety of targeted attacks
on individuals because the highest-ranked reconstructed records have a high chance of appearing in the original data.
As a conclusion, they raise ``sober warnings on the privacy risks of releasing precise aggregate statistics of a dataset''
and note that ``the only defenses against [reconstruction attacks] are to introduce imprecision in the underlying statistics
themselves, as techniques like differential privacy do''.
This is precisely what the USCB did in the 2020 Census release \cite{keller2023}.  

Despite the issues of the 2020 Census protection, CRR renewed the interest in using DP in census releases.
Specifically, in a subsequent article entitled ``Database Reconstruction Does Compromise Confidentiality'',
\cite{keller2023} ---by the current Chief Scientist at the USCB and her immediate predecessor--- 
fully endorsed Dick \emph{et al.}'s conclusions, both regarding the privacy threats of CRR,
and the advice to use DP to protect the released statistics.

Due to the (observed) adverse consequences of using DP in Census releases (or in any data release, since that was not the scenario DP was designed for \cite{Domingo2021,Blanco2023}), the flaws of previous reconstruction attacks, and the influence that the USCB's decisions may have on other statistical agencies or social science in general, the claims by Dick \emph{et al.} deserve detailed scrutiny. 

\subsection*{Contributions and plan}

In this paper, we empirically demonstrate the inability of the CRR attack to threaten privacy in any meaningful manner. In particular, we show that:
(i) the highest ranked records according to Dick \emph{et al.}'s confidence ranking are also the most common and, hence, the most inherently protected against reidentification;
(ii) rare or unique records, whose reconstruction might put the corresponding individuals at risk of disclosure, go unnoticed by the proposed confidence ranking; and
(iii) the inaccuracy of the reconstruction and the large diversity of the non-existent records it generates 
(records that do not exist in the original data) render attribute disclosure attacks ineffective.

The rest of the paper is organized as follows.
Section \ref{sec_reconstruction} reviews the CRR attack.
In Section \ref{sec_experiments} we detail how we replicated Dick \emph{et al.}'s experiments.
Sections \ref{sec_results} and \ref{sec_attribute} assess the actual reidentification and attribute disclosure risks implied by the CRR attack. Finally, we present some conclusions.

\section{Reviewing the confidence-ranked reconstruction attack}
\label{sec_reconstruction}

In and of itself, reconstruction poses no privacy risk unless \emph{it is accurate}. In the following, we demonstrate that the confidence-ranked reconstruction that Dick \emph{et al.} propose is ineffective both at accurately reconstructing Census records and at detecting, even approximately, records at risk.

The primary purpose of the CRR is to attach a confidence level to each reconstructed record that measures how likely it is for that record to appear also in the data they use as ground truth. This ground truth consists of synthetic microdata published by the USCB that closely resemble the real 2010 Census microdata.
By ``casting the reconstruction problem as an instance of large-scale, non-convex optimization, along with a subsequent step to convert non-continuous ({\em e.g.}, categorical) features back to their original schema'' \cite{dick2023confidence}, they rank reconstructed record prototypes by how frequently they appear in multiple reconstructions. 
A \emph{record prototype} is a record type with some \emph{multiplicity} (that is, the number of repetitions) in the microdata. We talk about prototypes rather than actual records because CRR is incapable of ascertaining the multiplicity of those prototypes and, therefore, \emph{it cannot produce an accurate reconstruction of the synthetic microdata}. 

Nevertheless, the authors interpret the rank of the reconstructed records as a measure of risk, because the empirical results they report show that the top $k$ ranked ({\em i.e.}, most frequent) reconstructed prototypes were present in the synthetic data set ($D$) in a large proportion. More specifically, they argue
that the most confident/highest ranked records they reconstruct are at risk because they are those that are most likely to appear in the original data and, therefore, can be the target of a variety of privacy attacks, including ``identity theft''. 

Dick \emph{et al.} measure the effectiveness of their CRR attack as the proportion of the top $k$ ranked prototypes ({\em i.e.}, the most frequently reconstructed ones) that were present in $D$.
According to their results, this proportion was near 1 for small values of $k$.
However, since the measured proportion considers the number of record prototypes in $D$ (without counting their repetitions), rather than the actual 
number of records (counting repetitions), the authors are neglecting the multiplicity of each prototype in $D$, which is key to privacy: a prototype appearing, say, 10 times in $D$ means that 10 individuals share the same record values for the considered attributes and, therefore, those individuals are intrinsically protected against reidentification (they are 10-anonymous, in terms of the $k$-anonymity privacy model \cite{Samarati2001}, and the probability of successfully reidentifying one of them is at most $1/10$). 

To better understand the importance of this aspect and to illustrate the practical ineffectiveness of the CCR attack, in the following, we exactly replicate the experiments done by Dick \emph{et al.} on the same data, and report the number of repetitions in $D$ of their reconstructed prototype (which the authors did not do).   

\section{Replicating Dick et al.'s experiments}
\label{sec_experiments}

Our experiments have been done using the code and settings provided by Dick \emph{et al.}\footnote{\url{https://github.com/terranceliu/rap-rank-reconstruction}}. 
The additional code we have written for our experiments is also available for reproducibility \footnote{\url{https://github.com/NajeebJebreel/CRR-analysis}}.
Specific details follow.

\textbf{Data Set:} We used the same subsets of synthetic U.S. Census microdata as ground truth. This dataset was released by the USCB and closely resembles the real 2010 Census microdata in terms of statistical characteristics. Specifically, we used the 2020-05-27 vintage Privacy-Protected Microdata File (PPMF) \cite{US20}, which consists of 312,471,327 rows representing synthetic responses for individuals in the 2010 Decennial Census. The columns in the PPMF include attributes such as the respondent's home location (state, county, census tract, and census block), housing type, sex, age, race, and Hispanic or Latino origin.

\textbf{Statistical Queries:} Dick \emph{et al.}'s reconstruction was executed on the tables employed by the USCB for their internal reconstruction attack on the 2010 Census data. Each table defines a set of statistical queries that are performed on the (protected) Census microdata. These queries involve specifying column names, subsets of column domains, and census block or tract identifiers. The objective is to count the number of microdata rows that satisfy the specified criteria.
The tables include P1 (total population), P6 (race), P7 (Hispanic or Latino origin by race), P9 (Hispanic or Latino and Not Hispanic or Latino by race), P11 (Hispanic or Latino and Not Hispanic or Latino by race for the population 18 years and over), P12 (sex by age for selected age categories), P12 A-I (sex by age for selected age categories iterated by race), PCT12 (sex by single year age), and PCT12 A-N (sex by single year age iterated by race). The P tables are released at the block level, while the PCT tables are released only at the census tract level.

\textbf{Experiments:} We used subsets of the PPMF, which comprised all the rows belonging to specific census tracts or blocks.
For tracts, we used the same random sample employed by Dick \emph{et al.} On the other hand, blocks were selected according to the following (non-random) criteria described by the authors~\cite{dick2023confidence}: for
each state, choose the block closest in size to the mean block size as well as the largest block; in addition, choose blocks closest in size to $M/C$, where $M$ is the maximum block size in the state and $C \in \{2, 4, 8, 16\}$.
This resulted in a total of 50 tracts and 300 blocks. 
Then, we reconstructed 100 data sets for each original data set, by employing both \emph{baseline} and \emph{random} initialization methods, while adhering to the same configuration and parameters. Baseline initialization follows the distribution of $D$ ---assumed to be publicly known--- to set up the algorithm's parameters, whereas random initialization leverages no prior information.

\section{Results and discussion}
\label{sec_results}

Having replicated Dick \emph{et al.}'s experiments, we investigated the correlation between the multiplicities of records in $D$ and their corresponding rank in the reconstructed data. Our aim was to determine whether highly ranked records threaten the privacy of individuals as Dick \emph{et al.} claim.

\begin{figure*}[!htbp]
    \centering
    \begin{subfigure}{0.5\textwidth}
      \centering
      \includegraphics[width=1\linewidth]{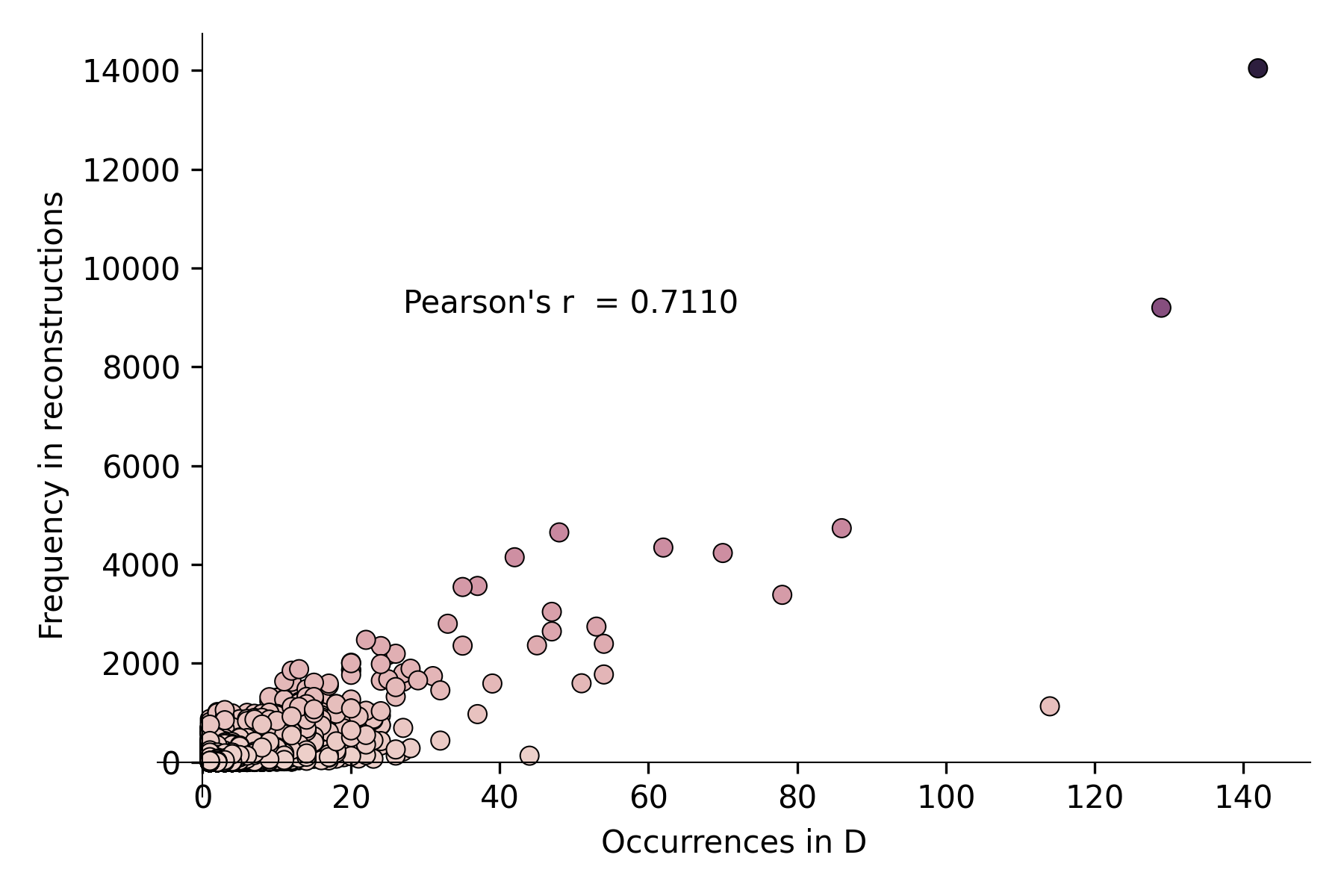}
      \caption{Tract (random)}
      \label{fig:corr_tract_random}
    \end{subfigure}% 
    \begin{subfigure}{0.5\textwidth}
      \centering
      \includegraphics[width=1\linewidth]{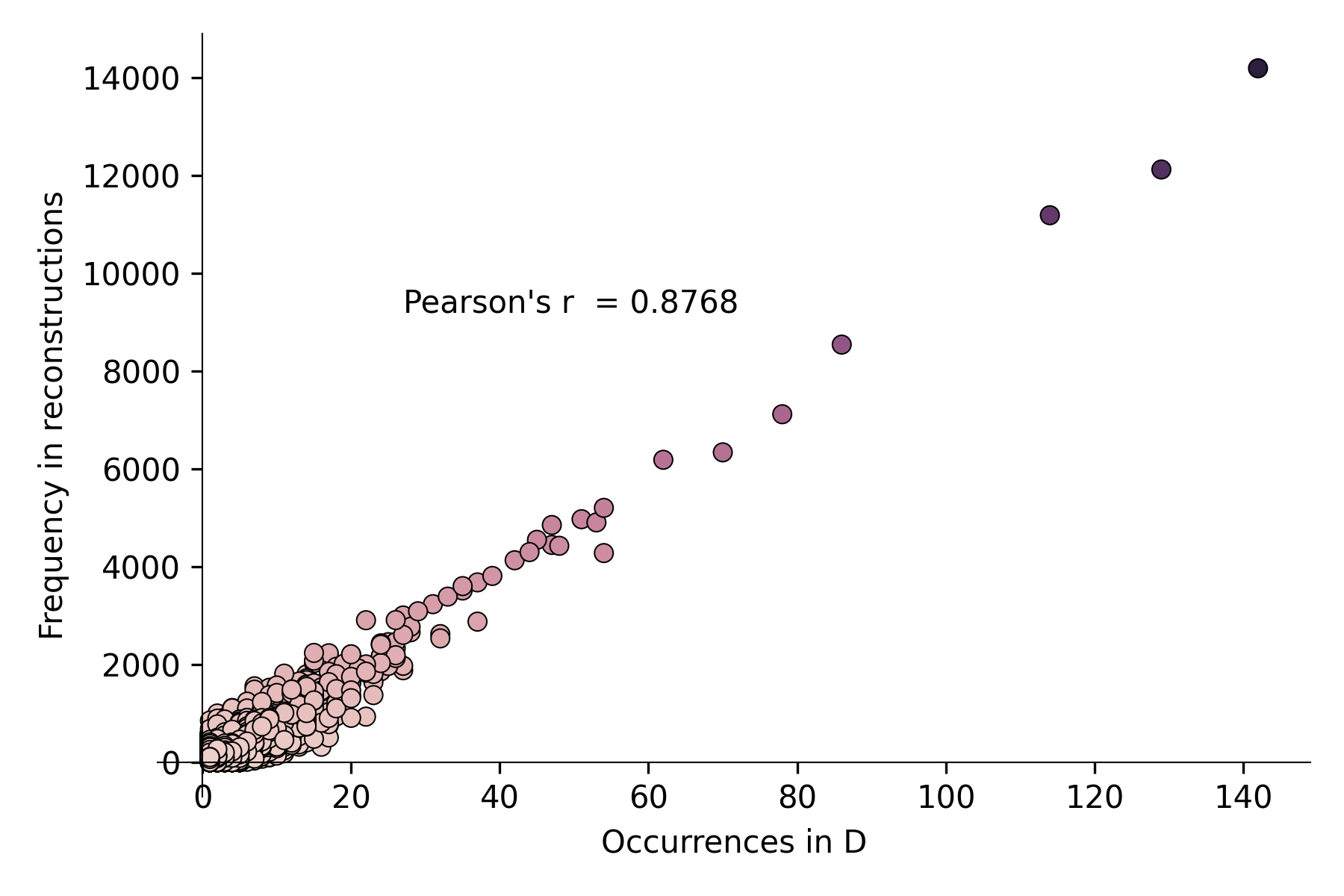}
      \caption{Tract (baseline)}
      \label{fig:corr_tract_baseline}
    \end{subfigure}

    \begin{subfigure}{0.5\textwidth}
      \centering
      \includegraphics[width=1\linewidth]{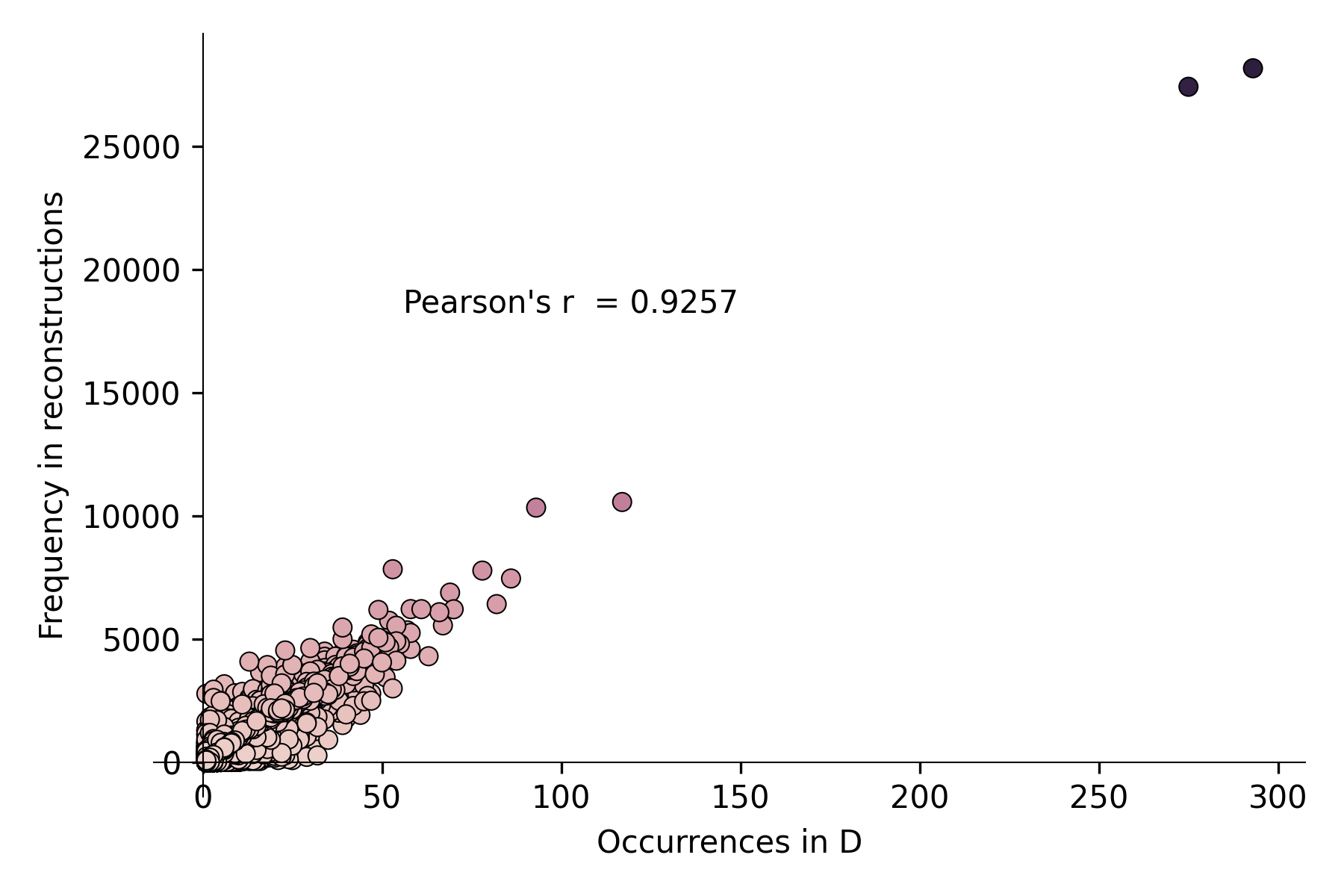}
      \caption{Tract without block (random)}
      \label{fig:corr_tract_block_random}
    \end{subfigure}% 
    
    \begin{subfigure}{0.5\textwidth}
      \centering
      \includegraphics[width=1\linewidth]{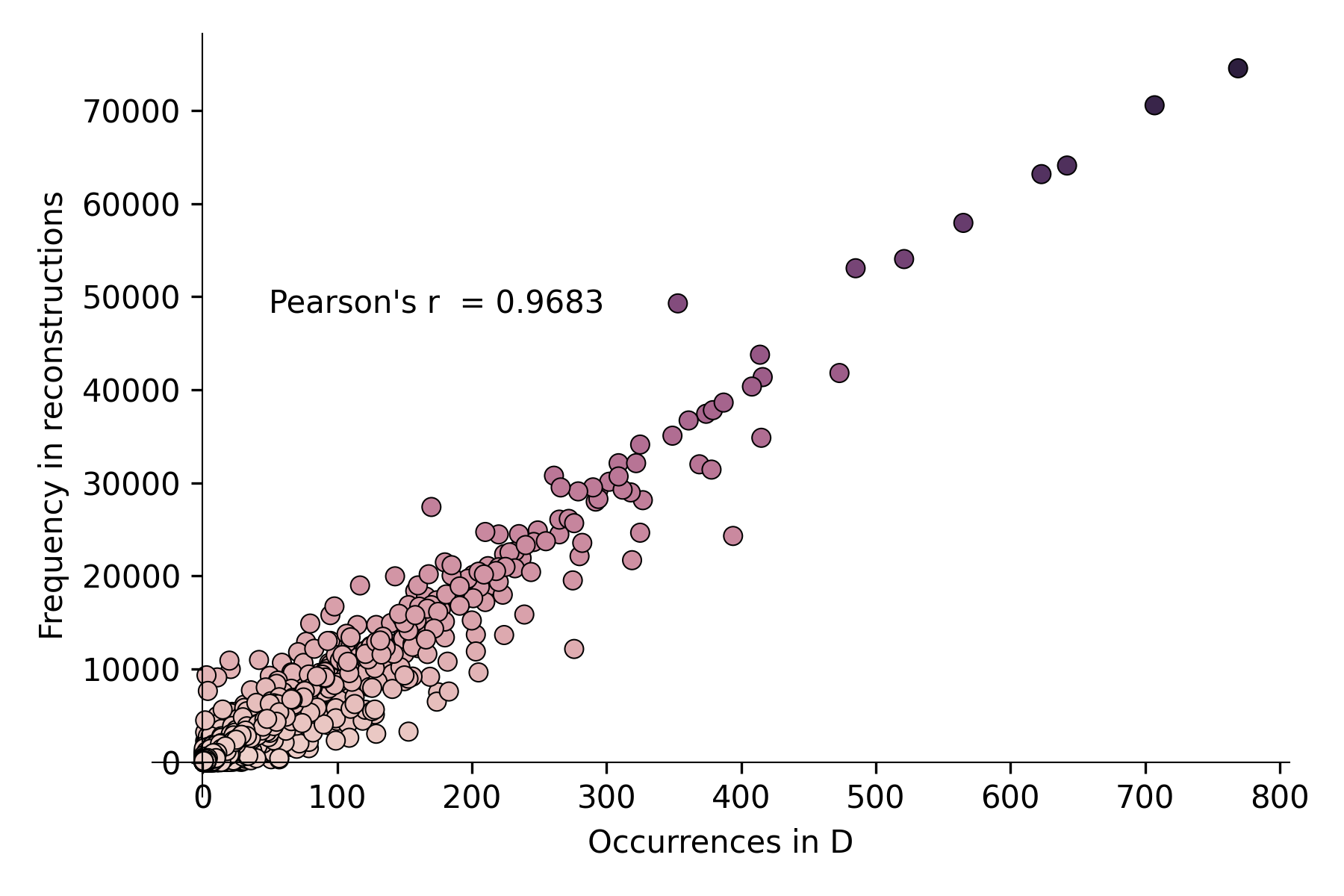}
      \caption{Block (random)}
      \label{fig:corr_block_random}
    \end{subfigure}%
    \begin{subfigure}{0.5\textwidth}
      \centering
      \includegraphics[width=1\linewidth]{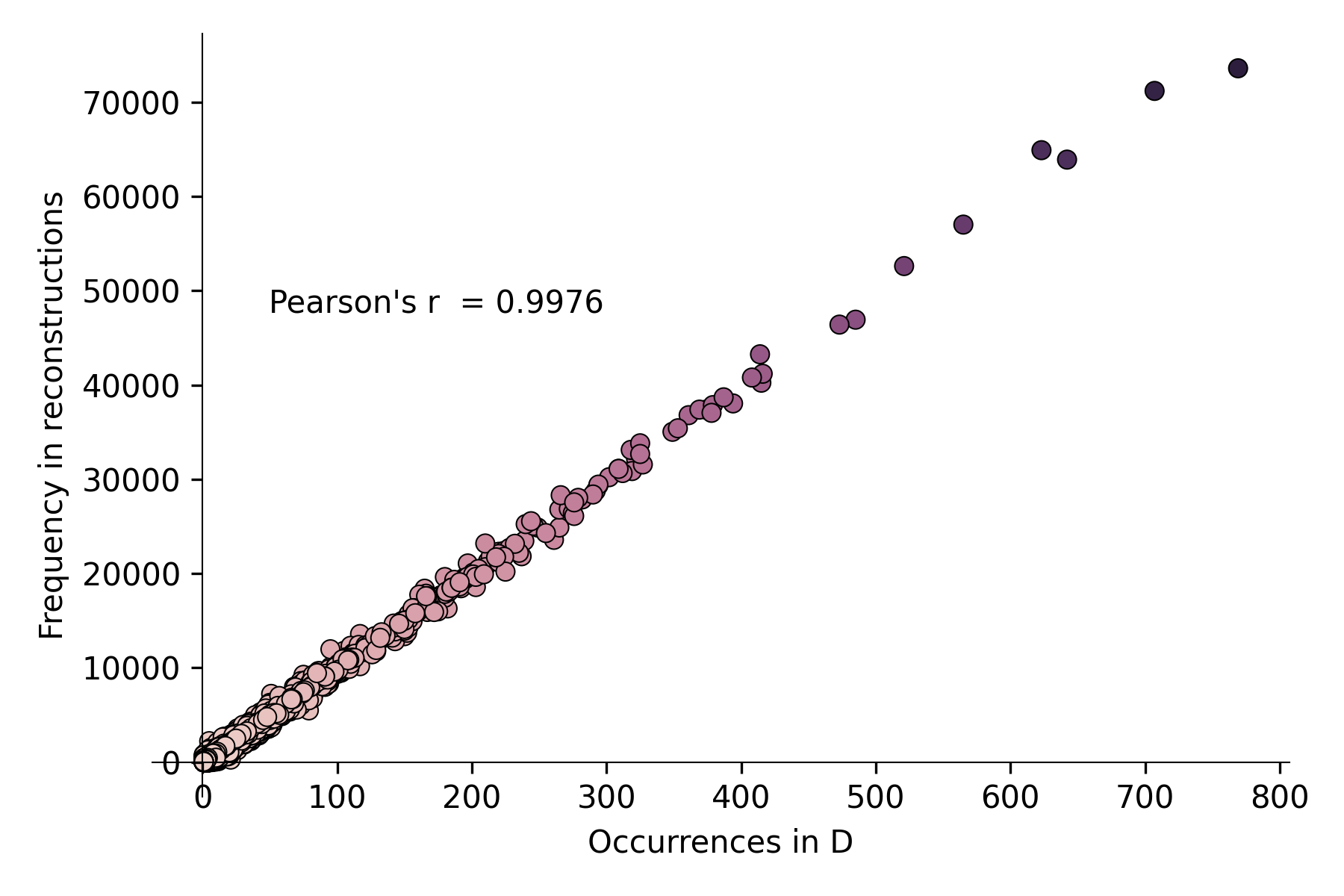}
      \caption{Block (baseline)}
      \label{fig:corr_block_baseline}
    \end{subfigure}%
\caption{Comparison between the multiplicities of record prototypes in $D$ and their frequency in CRR.}
\label{fig:census_correlation}
\end{figure*}

To do so, we counted the number of occurrences of each record prototype in $D$ and we compared it with its frequency in the reconstructions (which is the foundation of the confidence ranking).
Figure~\ref{fig:census_correlation} reports this comparison both at the tract and block levels and both with random and baseline initializations. We also report results for tracts without considering the block attribute (random initialization), which Dick \emph{et al.} removed to ease tract reconstructions.

Our results clearly show that records with larger multiplicity in $D$ have a higher frequency (rank) in the CRR. This dependency is very strong in all cases (Pearson's correlation $r$ between 0.71 and 0.99), but especially 
for block random or baseline initialization (where $r>0.96$). As a matter of fact, in all cases, the record with the largest multiplicity in $D$ was also the most frequent/top-ranked prototype in the CRR. 
Since the confidence ranking is based on the frequency of record prototypes in multiple reconstructions, the top-ranked prototypes are the $k$ most common prototypes in $D$; that is, those with the greatest number of repetitions and, thus, those intrinsically most private. 
Therefore, all records top-ranked by CRR ({\em i.e.}, those claimed to be at risk) are intrinsically protected against reidentification because ``our privacy is protected to the extent that we blend in with the crowd'' \cite{gehrke2012}. 

Conversely, rare or unique records in $D$, which are actually vulnerable to reidentification because of their rarity, do not appear among the top-ranked $k$. Hence, records really at risk stay paradoxically undetected by CRR.

By examining the distribution of $D$, we see that the percentage of unique records is 10.17\% at the block level and 1.85\% at the tract level. Hence, 89.83\% are non-unique at the block level and 98.15\% are non-unique at the tract level. 
Also, the multiplicity of the most common records is in the order of dozens at the tract level and in the order of hundreds at the block level. 
In terms of the well-known $k$-anonymity model \cite{Samarati2001}, this means that unequivocal reidentification of any of those 89.83\% and 98.15\% records is not possible and that the probability of correct
reidentification is as low as 1/140 at tract level (1/300 when dropping the block attribute), and 1/800 at the block level. From these results, it can be seen that CRR would still (wrongly) claim high rates if $D$ was already protected via $k$-anonymity, {\em i.e.}, all prototypes in 
it had multiplicity $k>1$. Therefore, it is clear that CRR does not capture the reidentification risk in $D$. 

Notice that, since tracts are subdivided into blocks, one would expect the multiplicity in tracts to be larger than in blocks. However, the figures above show the opposite. These counterintuitive results are because we exactly replicated Dick \emph{et al.}'s data sampling, where the sampling of blocks is independent from the sampling of tracts (both are done at the state level). More importantly, whereas tracts were randomly sampled, the largest and mean blocks per state were deterministically chosen. This yields blocks that tend to be larger than tracts. No reason is given in \cite{dick2023confidence} for the different (non-random) criteria they employed to choose blocks.
One motivation for choosing large blocks is that block-level data become more homogeneous because large blocks correspond to dense populations that tend to have inhabitants with similar profiles. This results in a greater amount of common records, which are those the confidence ranking scores higher.

Given the strong positive correlation between the multiplicity of prototypes and their 
confidence ranking, one might argue that the lowest-ranked record prototypes may be used to detect unique records in $D$ and, therefore, indicate a privacy threat. In the following, we argue that this is not the case because of two reasons.

First, we must consider how Census data were protected in the 2010 release, on which all reconstruction attacks in the literature are based. Protection in that release was performed via \emph{data swapping}. According to the USCB \cite{zayatz2009}, the selection of the records to be swapped was highly targeted to the records with the most disclosure risk, that is, those that were unique in their block based on a set of key demographic variables. Also, the probability of being swapped had an inverse relationship with block size. Data from these households at risk were swapped with data from other households that had similar characteristics on a certain set of variables but were from different geographic locations. 

According to the description above, the swapping methodology was targeted toward certain records that were defined as at risk of disclosure ({\em i.e.}, vulnerable records). Even though the exact details on how vulnerable records were determined are not public, the selection criterion was based on their uniqueness in their block \cite{swapping}.
Given that the statistics employed for CRR come from the swapped/protected data (and not the \emph{original} data), we can deduce that reconstructed unique records are more likely to be protected (that is, swapped) records than original records. 

Second, in order to be able to reconstruct (even protected) unique records in $D$, this should be done exhaustively and unequivocally. For this to be possible with CRR, records that are rare in the reconstructed data should also appear as rare in $D$ with high certainty.
To see how likely this is to happen, we calculated the following proportions:
\begin{itemize}
\item number of prototypes that occur once in $D$ and occur once in the reconstructions over the total number of prototypes that occur once in the reconstructions;
\item number of prototypes that occur twice in $D$ and occur once or twice in the reconstructions over the total number of prototypes that occur once or twice in the reconstructions;
\item number of prototypes that occur three times in $D$ and occur once, twice, or three times in the reconstructions over the total number of prototypes that occur once, twice, or three times in the reconstructions.
\end{itemize}

Table~\ref{tab:percentage_vulnerable} reports the above proportions as percentages, and they are extremely small. 
In other words, the certainty that a rare reconstructed prototype occurs in $D$ is almost zero.

\begin{table*}[ht!]
\centering
\small
\caption{Percentage of rare record prototypes in reconstructions that occur once, twice, or three times in $D$. Results are given for tracts, for tracts without block, and for blocks.}
\label{tab:percentage_vulnerable}
\resizebox{\textwidth}{!}{
\begin{tabular}{lccccccccc} \toprule
Level                                 & \multicolumn{3}{c}{Tract} & \multicolumn{3}{c}{Tract (w/o block)} & \multicolumn{3}{c}{Block}  \\ \midrule
Occur. $D$/ freq. CRR & 1/1  & 2/$\leq$2 & 3/$\leq$3         & 1/1  & 2/$\leq$2 & 3/$\leq$3 & 1/1  & 2/$\leq$2 & 3/$\leq$3         \\ \midrule
Random                                & 0.09\% & 0.08\% & 0.05\%         & 0.15\% & 0.16\% & 0.15\%         & 0.20\% & 0.15\% & 0.12\%         \\
Baseline                              & 0.09\% & 0.04\% & 0.02\%         & N/A & N/A & N/A         & 0.15\% & 0.06\% & 0.03\%         \\ \bottomrule
\end{tabular}}
\end{table*}

To further illustrate the ineffectiveness of CRR as a reconstruction method, in Table \ref{tab:percetage_appearance} we report the percentage of reconstructed prototypes that did not occur in $D$ (w.r.t. the total number of record prototypes in $D$).

\begin{table}[ht!]
\centering
\small
\caption{Percentage of reconstructed record prototypes that did not occur in $D$}
\label{tab:percetage_appearance}
\begin{tabular}{lccc} \toprule
Level        & Tract & Tract (w/o block) & Block  \\ \midrule
Random          & 4805.3\% & 703.2\% & 820.1\%     \\
Baseline        & 2223.6\%  & N/A &  449.9\%     \\ \bottomrule
\end{tabular}
\end{table}

We can see that Dick \emph{et al.}'s ``optimization-based'' method generates a very large number of non-existent record prototypes, which is 4 to 48 times larger than the number of prototypes in $D$. As discussed above, this results in very large uncertainty when reconstructing the rarest records in $D$, because there is a very high probability that the reconstructed prototypes do not exist in the original data. 

On the other hand, by counting the percentage of records in $D$ that did not appear in any of the reconstructions, we obtain the non-negligible figures reported in Table \ref{tab:percetage_non}.

\begin{table}[ht!]
\centering
\small
\caption{Percentage of record prototypes in $D$ that did not appear in the reconstructions}
\label{tab:percetage_non}
\begin{tabular}{lccc} \toprule
Level        & Tract & Tract (w/o block) & Block  \\ \midrule
Random          & 8.16\% & 1.97\% & 3.31\%     \\
Baseline        & 3.07\%  & N/A & 1.01\%     \\ \bottomrule
\end{tabular}
\end{table}

\section{On attribute disclosure risk}
\label{sec_attribute}

In a yet more recent article \cite{dick2023reply}, the authors argued that the main risk posed by CRR is actually \emph{attribute disclosure}, that is, unequivocal inference by the adversary of confidential attribute values of known individuals, even without being able to reidentify them in the data set. This threat was not mentioned in their initial paper describing CRR \cite{dick2023confidence}, which only discussed privacy attacks related to identity disclosure. In that new paper, the authors fell short of supporting this new claim with any theoretical or empirical evidence. They just conjectured attribute disclosure through the following single fabricated example referred to a different dataset (the American Community Survey (ACS)):

\begin{quote}
Suppose that it is known that there are two 46-year-old married men with a particular racial designation and level of educational attainment within the [ACS] dataset. These features might be publicly known since they are not features the individuals intend to hide. Even if these two individuals match on all other features as well ---that 
is, they share the same record prototype--- if we are able to learn it, then we have learned facts about both of them 
({\em e.g.}, their citizenship status, their income, etc.) that they may not have wanted to share. It does not matter that we cannot determine which record corresponds to which individual (the question does not even make sense, as the records are identical) since we have learned private information about both.
\end{quote}

For the described privacy threat to lead to unequivocal inference of confidential attribute values on the known individuals, it is not enough to accurately generate the corresponding record prototype. The following must also hold: 
\begin{enumerate}
    \item The original data set must not contain any other records sharing the same values for the public attributes with the known individuals (in the above quotation, age, marital status, race, and education) but having different values for the confidential attributes (in the above quotation, citizenship status and income).
    \item The original data set should be an exhaustive sample of the population to which the known individuals belong. Otherwise, we cannot be sure whether the known individuals are or are not present in the data set and, therefore, whether the potentially reconstructed records matching their public attributes correspond to them. Even though this holds for the Decennial Census (as long as the known individuals live in the U.S.), it is not the case for other non-exhaustive datasets, such as the American Community Survey (ACS) that Dick \emph{et al.} used to illustrate the attribute disclosure threat. 
    \item CRR must not generate record prototypes sharing the same values for the public attributes but having diverse values for the confidential attributes.
\end{enumerate}

Even though 1) and 2) may hold in practice, we have shown in Section \ref{sec_results} above that CRR was not only inaccurate but that it generated an enormous diversity of non-existent records (see Table \ref{tab:percetage_appearance}). Therefore, 3) will not hold, and the diversity of the (erroneous) confidential values in the reconstructed records will prevent unequivocal attribute disclosure. In fact, it is well documented that diversity of confidential attributes effectively prevents attribute disclosure in $k$-anonymous-like data releases~\cite{ldiversity,tcloseness}.

\section{Conclusions}
\label{sec_conclusions}

We have shown that, despite the claims of~\cite{dick2023confidence} and~\cite{dick2023reply}, the proposed CRR attack is ineffective to
i) detect the most privacy-sensitive records,
ii) guide targeted reidentification attacks, or
iii) guide attribute disclosure attacks. 
Moreover, we have shown that the ``optimization-based'' method by Dick \emph{et al.} fails to reconstruct a significant proportion of original records while generating an enormous amount of records that do not exist in
the original data set.
The latter adds a very large uncertainty to disclosure inferences, whether aimed at reidentification or attribute disclosure.
Having demonstrated that the CRR attack implies no privacy risks, we can conclude that it cannot be used to justify 
the use of (utility-damaging) DP methods ---contrary to what \cite{dick2023confidence} and \cite{abowd2023} claim.

\section*{Acknowledgments}

This research was funded by the the European Commission (project H2020-871042 ``SoBigData++''), the Government of Catalonia (ICREA Acad\`emia Prizes to J. Domingo-Ferrer and to D. S\'anchez), MCIN/AEI/ 10.13039/501100011033 and ``ERDF A way of making Europe'' under grants PID2021-123637NB-I00 ``CURLING'' and {PRE2019-089210}, and INCIBE and European Union NextGenerationEU/PRTR (project ``HERMES'' and INCIBE-URV Cybersecurity Chair).

\bibliographystyle{splncs04}
\bibliography{References}

\end{document}